\documentclass[conference]{IEEEtran}
\IEEEoverridecommandlockouts
\usepackage{graphicx}          
\usepackage{amsmath}
\usepackage{tabularx}
\usepackage{amsfonts}
\usepackage{url}
\usepackage{verbatim}
\usepackage{multirow}
\usepackage{times,epsfig}
\usepackage{makecell}
\usepackage{psfrag}
\usepackage{subfigure}
\usepackage{stfloats}
\usepackage{caption}
\usepackage{makecell}
\usepackage{setspace}
\usepackage{diagbox}
\usepackage{amssymb}
\usepackage[short]{newoptidef}
\usepackage{multirow}
\usepackage{colortbl}
\usepackage[justification=centering]{caption}
\usepackage{fancyhdr}
\usepackage[table,xcdraw]{xcolor}

\begin{document}
	
	\pagenumbering{arabic}
	
  \title{Joint Foundation Model Caching and Inference of Generative AI Services for Edge Intelligence}
\author{\IEEEauthorblockN{Minrui Xu$^{\mathrm{1}}$, Dusit Niyato$^{\mathrm{1}}$, Hongliang Zhang$^{\mathrm{2}}$, Jiawen Kang$^{\mathrm{3}}$, Zehui Xiong$^{\mathrm{4}}$, Shiwen Mao$^{\mathrm{5}}$, and Zhu Han$^{\mathrm{6,7}}$}
\IEEEauthorblockA{$^{\mathrm{1}}$School of Computer Science and Engineering, Nanyang Technological University, Singapore 639798, Singapore \\
$^{\mathrm{2}}$School of Electronics, Peking University, Beijing 100871, China \\
$^{\mathrm{3}}$School of Automation, Guangdong University of Technology, Guangzhou 510006, China \\
$^{\mathrm{4}}$Singapore University of Technology and Design, Singapore 487372, Singapore \\
$^{\mathrm{5}}$Department of Electrical and Computer Engineering, Auburn University, Auburn, AL 36849-5201, USA \\
$^{\mathrm{6}}$Department of Electrical and Computer Engineering, University of Houston, Houston, TX 77004, USA \\
$^{\mathrm{7}}$Department of Computer Science and Engineering, Kyung Hee University, Seoul 446-701, South Korea \\}
Email: minrui001@e.ntu.edu.sg, dniyato@ntu.edu.sg, hongliang.zhang@pku.edu.cn, kavinkang@gdut.edu.cn, \\zehui\_xiong@sutd.edu.sg, smao@ieee.org, hanzhu22@gmail.com.\\
}
\maketitle
\pagestyle{headings}

\begin{abstract}
With the rapid development of artificial general intelligence (AGI), various multimedia services based on pretrained foundation models (PFMs) need to be effectively deployed. With edge servers that have cloud-level computing power, edge intelligence can extend the capabilities of AGI to mobile edge networks. However, compared with cloud data centers, resource-limited edge servers can only cache and execute a small number of PFMs, which typically consist of billions of parameters and require intensive computing power and GPU memory during inference. To address this challenge, in this paper, we propose a joint foundation model caching and inference framework that aims to balance the tradeoff among inference latency, accuracy, and resource consumption by managing cached PFMs and user requests efficiently during the provisioning of generative AI services. Specifically, considering the in-context learning ability of PFMs, a new metric named the \emph{Age of Context (AoC)}, is proposed to model the freshness and relevance between examples in past demonstrations and current service requests. Based on the AoC, we propose a least context caching algorithm to manage cached PFMs at edge servers with historical prompts and inference results. The numerical results demonstrate that the proposed algorithm can reduce system costs compared with existing baselines by effectively utilizing contextual information.
\end{abstract}

\section{Introduction}
Moving towards Artificial General Intelligence (AGI) in mobile edge networks~\cite{bubeck2023sparks, zhou2022vetaverse}, pre-trained foundation models (PFMs), such as generative pre-trained transformers (GPTs)~\cite{brown2020language}, have achieved great successes in a variety of fields over the past few years. As building blocks of AGI, PFMs with billions of parameters are essential due to their effectiveness in demonstrating emergent capabilities in downstream tasks with various data modalities~\cite{zhou2023comprehensive}. The pre-training approach provides an efficient parameter initialization for a wide range of downstream tasks, including semantic segmentation, content generation, and information retrieval. As a result, language/visual/multimodal foundation models belong to the paradigm of transfer learning, which can adapt to new tasks and domains without any task-specific data during pre-training. 

Multimedia services based on edge intelligence, such as intelligent digital twins (DTs), autonomous driving, and AI-generated content (AIGC), can be greatly enhanced by deploying PFMs on edge servers, benefiting from edge computing's low latency and flexible features. For instance, in autonomous driving, PFMs can generate traffic simulations and provide driving assistance in making complex driving decisions~\cite{xu2023generative}. 
Additionally, during immersive human-avatar interactions in the Metaverse, PFMs can assist in comprehending and reacting to human emotions and behaviors. For example, ChatGPT facilitates consistent and fluent interactions with humans, fine-tuned based on GPT-3 to release its contextual awareness~\cite{brown2020language}, which is an LFM with 175 billion parameters. Beyond executing PFMs in cloud data centers, edge servers can support fine-tuning and inference processes of PFMs requested by AI services, thus igniting the sparks of AGI in mobile edge networks. 

However, unlike cloud data centers, resource-constrained edge servers are unable to concurrently load all PFMs to serve users' AI service requests. In literature, existing research generally focuses on offloading AI services to cloud data centers for remote execution or caching inference results at edge servers for low-latency response~\cite{gilman2019challenges}. On one hand, offloading inference requests of PFMs to cloud data centers introduce additional latency, traffic overhead, and privacy threats to serving AI services over core networks and public cloud infrastructure. On the other hand, merely caching inference results at edge servers is no longer effective for satisfying users' interactive requirements. To enable mobile AI services with the computing and GPU resources currently loaded into the GPUs of edge servers, effective deployment of PFMs at edge servers requires flexible and context-aware management on computing resources and user requests.


Differing from the existing works on joint service caching and task offloading, several unique challenges arise for joint foundation model caching and inference to balance the tradeoff among inference latency, accuracy, and resource consumption in mobile edge networks~\cite{zhou2019edge}. First, different quantities of requests and performance requirements of the downstream tasks, such as accuracy and latency, are present during the fine-tuning and inference of PFMs~\cite{gilman2019challenges}. Additionally, a variety of PFMs can be applied to comparable downstream tasks in a range of AI services. This presents a challenge for edge servers in that the cached PFMs may be called interchangeably to handle model misses. Furthermore, PFMs can continuously learn and adapt to new domains and tasks through prompts of instruction and interactive demonstrations~\cite{dong2022survey}. Due to the in-context learning ability of PFMs, cached models can enhance their inference accuracy during inference without parameter updates. These challenges make decisions about cached model management and request offloading increasingly difficult for optimizing the performance of the framework, which is a tradeoff among inference latency, accuracy, and resource consumption.

To address these issues, in this paper, we investigate the important but rarely studied problem of joint foundation model caching and inference of generative AI services for edge intelligence in mobile edge networks. We propose a joint foundation model caching and inference framework to serve PFMs for provisioning generative AI services.
Furthermore, to balance the tradeoff among inference latency, accuracy, and resource consumption, we propose a new metric named \emph{Age of Context} (AoC) to indicate the freshness and relevance between examples in historical demonstrations and current inference requests. With a context vanishing factor, the AoC follows the non-increasing utility function that affects the effective examples in context from instruction, demonstrations, and outputs of past interactions. Based on the AoC, we propose a Least Context (LC) algorithm to manage cached PFMs at edge servers. Simulation experiments demonstrate that the proposed LC algorithm can reduce the total system cost by utilizing contextual information for improving the service accuracy and utilizing the computing power and GPU memory of edge servers efficiently.

The main contributions of this paper are summarized as follows.
\begin{itemize}
    \item For the first time, we formulate the joint foundation model caching and inference problem in mobile edge networks, for minimizing service cost and accuracy loss under limited computing and GPU memory capacity of edge servers.
    \item Considering the in-context learning ability of PFMs, we propose a new metric named age of context to measure the freshness and relevance of historical examples in context and current inference requests.
    \item Based on the AoC, we develop the least context algorithm to efficiently manage the cached models by utilizing the contextual information and thus reducing model switching, inference, and accuracy costs.
\end{itemize}
Compared with our prior work in~\cite{xu2023sparks}, this paper provides formal mathematical formulations for the joint foundation model caching and inference problem, the new age of context metric, and the least context algorithm.

\begin{figure}[t]
    \centering
    \includegraphics[width=0.93\linewidth]{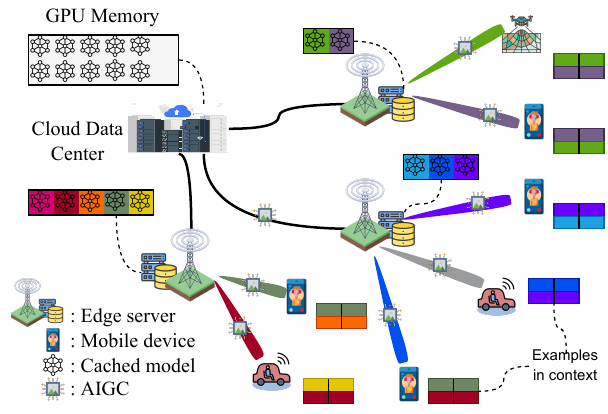}
    \caption{Joint foundation model caching and inference of generative AI services for edge intelligence.}
    \label{fig:systme}
\end{figure}
\section{System Model}

As shown in Fig.~\ref{fig:systme}, we consider an edge intelligence system model consisting of service providers, including one cloud data center and a set of edge servers, and a massive amount of users. The cloud data center and edge servers can serve  generative AI services. 
The cloud data center is represented by $0$ and the set of edge servers is represented by $\mathcal{N} = \{1, 2, \ldots, N\}$. In this system, edge servers and the cloud center provide generic AI services such as AIGC, depending on different PFMs. We use a set $\mathcal{I}=\{1,2,\ldots, I\}$ to denote the available generative AI services based on a set of PFMs $\mathcal{M}=\{1, 2, \ldots, M\}$. As PFMs are capable of performing multiple downstream tasks in generative AI services simultaneously, and thus we consider $I \gg M$ that the number of AI servers is far greater than the number of PFMs.

Mobile users must request generative AI services from edge servers or cloud data centers when their devices are insufficient for executing PFMs. The inference requests of the diverse services might request different PFMs when they are serving different functions. Typically, a generative AI service requires the collaboration of several PFMs to process users' requests. For instance, in Stable Diffusion~\cite{rombach2022high}, text-related conditioning is based on a pre-trained CLIP ViT-L/14 model. Then, a variational autoencoder compresses images into a smaller dimensional latent space. Finally, a U-Net block is used to denoise the output from forward diffusion backwardly to obtain a latent representation. This is a typical process of serving text-to-image generation service requests. We use $R_{n, i,m}^t$ to denote the number of inference requests generated by AI service $i$ to execute foundation model $m$ at edge server $n$. The configuration of PFM $m$ consists of the amount of runtime GPU memory, which is proportion to model size $s_m$, the inference cost per token $e_m$, the model accuracy $a_m$, and the size of context window $w_m$. The inference process of AI services can put certain context information into the context window of models. Then, the number of examples in context is denoted by $K^t_{i,m}$ of model $m$ for application $i$ which is zero initially, i.e., $K^0_{i,m}=0$.

\subsection{Decision Variables}

To offer AI services based on PFMs, we propose a joint foundation model caching and inference framework. Edge servers need to make model caching and request offloading decisions to utilize the existing edge computing resources for accommodating generative AI service requests of mobile users. Specifically, edge server $n$ needs to determine the following variables: (i) Let $a_{n, i, m}^t \in \{0,1\}$ denote the binary variable indicating whether model $m$ of application $i$ is cached at edge server $n$ at time slot $t$; (ii) Let $b_{n, i, m}^t \in [0,1]$ denote the continuous variable on whether model $m$ of application $i$ is cached at edge server $n$ at time slot $t$. Let $\mathbf{a}^t_n = \{a^t_{n,1,1}, \ldots, a^t_{n,I,M}\}$ denote the model caching decisions of edge server $n$ and $\mathbf{a}^t = \{\mathbf{a}^t_1, \ldots, \mathbf{a}^t_n\}$. In addition, the request offloading decision of edge server $n$ can be denoted as $\mathbf{b}^t_n = \{b^t_{n,1,1}, \ldots, b^t_{n, I, M}\}$ and the request offloading decisions of all edge servers can be denoted as $\mathbf{b}^t = \{\mathbf{b}^t_1, \ldots, \mathbf{b}^t_n\}$.

The generative AI service requests of users can be executed at edge servers if the required components of models are loaded at the GPU memories. Let $G_n$ denote the capacity of GPU memory of edge server $n$. Then, the model caching decision variables are subjected to the following constraint
\begin{equation}
    \sum_{i\in \mathcal{I}} \sum_{m \in \mathcal{M}} a_{n, i, m}^t s_{m} \leq G_n, \forall n\in \mathcal{N}.
\end{equation}
The models of AI services can be executed at the edge server after they are loaded into the GPU memory. Therefore, the constraint of model execution at edge servers is
\begin{equation}
    b_{n, i, m}^t \mathbf{1}(R_{n,i,m}^t>0) \leq a_{n, i, m}^t, \forall n\in \mathcal{N}, i\in\mathcal{I}, m\in\mathcal{M}
\end{equation}
at time slot $t$ and $\mathbf{1}(\cdot)$ is the indicator function. Let $E_n$ denote the resource capacity of edge server $n$. The total resource consumption of servers is constrained by the total energy capacity, which can be represented as
\begin{equation}
     \sum_{i\in \mathcal{I}} \sum_{m \in \mathcal{M}} e_m a_{n, i, m}^t b_{n, i, m}^t R_{n, i,m}^t \leq E_n, n\in \mathcal{N}.
\end{equation}
In cloud data centers, there are no GPU memory constraints or energy constraints for cached PFMs.

\subsection{Age of Context and In-context Learning Accuracy}

PFMs, such as GPT-3, have the ability to perform in-context learning, which means that they can learn from past prompts and inference results when an unseen task is presented to them. Some primary experiments show that larger models are more effective at using in-context instructions and demonstrations, as demonstrated by their improved ability to learn a task from contextual information~\cite{brown2020language}. This is particularly useful in NLP tasks, where understanding the context of a sentence or paragraph is crucial for accurate interpretation. Based on the evidence that GPT-3 is capable of in-context learning, which contributes to its strong performance on a variety of language tasks, such as translation, basic arithmetic, and Q\&A. Let $K_{i,m}^t$ denote the number of effective examples of model $m$ for application $i$. The examples in the demonstration might have different impacts on the model performance in terms of relevance, quality, and freshness. We propose the AoC to measure the freshness of examples in demonstrations that have an impact on the quality of services provided by PFMs in tasks that are now being carried out downstream. For instance, the historical Q\&A records that are recorded during PFM inference can be used to improve future inference accuracy. These examples can be used to increase the accuracy of PFMs, as PFMs can use meta-gradient learn during interaction to fit them~\cite {dai2022can}. However, depending on the caliber, applicability, and timeliness of examples, the meta-gradient may have favorable or unfavorable impacts on the model performance. Similar to the definition of age of information (AoI), the AoC measures the freshness of historical contextual examples in demonstrations between the cached PFMs and the inference requests. As shown in TABLE~\ref{tab:accuracy},  with a vanishing factor $\nu_{i,m}$ of context, the AoC is adjusted by the non-increasing age utility function. Therefore, the effective number of examples in context $K_{i,m}^t$ at edge server $n$ can be represented as
\begin{equation}
    K_{i,m}^t = \min\left(w_m, \{K_{i,m}^{t-1} +  R_{n,i,m}^t a_{n,i,m}^{t} b_{n,i,m}^{t} - \nu_{i,m}\}^+\right),
\end{equation}
for $t = 1, \ldots, T$.
According to the AoC, the weighted total of the number of examples in demonstrations may be used to determine the number of examples in context.

As shown in Table~\ref{tab:accuracy}, the in-context (few-show) accuracy $A_{i,m}$ of model $m$ for the downstream task in application $i$ can be fit by a logarithmic function as~\cite{brown2020language}
\begin{equation}
    A_{i,m}(K_{i,m}^t) = A^0_m + A^1_m \log_2(1+{K_{i,m}^t}^{\alpha_m}),
\end{equation}
where $A^0_m$ is the zero-shot accuracy, $A^1_m$ is the one-shot accuracy, $K_{i,m}^t$ is the number of examples in context, and $\alpha_m$ is the coefficient of model $m$. 

\begin{table}[t]
\vspace{0.2cm}
\small\centering
\caption{The Parameters of Accuracy in Downstream Tasks of GPT3-13B/175B~\cite{brown2020language}.}
\label{tab:accuracy}
\begin{tabular}{|c|c|c|c|c|c|}
\hline
Task                              & Model & $K$  & $A^0$     & $A^1$     & $\alpha$      \\ \hline
\multirow{2}{*}{Translation}      & 13B   & 64 & 15.45 & 11.8  & 0.0923   \\ \cline{2-6} 
                                  & 175B  & 64 & 22.03 & 7.59  & 0.1565   \\ \hline
\multirow{2}{*}{Basic Arithmetic} & 13B   & 50 & 3.79  & 12.19 & -0.0501 \\ \cline{2-6} 
                                  & 175B  & 50 & 25.99 & 14.72 & 0.1813   \\ \hline
\multirow{2}{*}{SuperGLUE}        & 13B   & 32 & 54.40 & 9.89  & 0.0969  \\ \cline{2-6} 
                                  & 175B  & 32 & 58.20 & 10.70 & 0.1431   \\ \hline
\end{tabular}
\end{table}


\subsection{Cost Structure}

As discussed above, the generative AI service requests can be executed by edge servers and offloaded to cloud data centers over the core network. Given the model caching and request offloading decisions, the total system cost of serving generative AI services consisting of the edge inference cost and cloud inference cost can be formulated as follows.

\subsubsection{Edge Inference Cost}
Specifically, the edge inference cost consists of the edge switching cost, the edge transmission cost, the edge computing cost, and the model accuracy cost.
According to model caching decisions, each edge server needs to load models into the GPU memory before execution. During the loading process, the model switching cost consisting of the model loading latency and hardware wear-and-tear cost is incurred. Therefore, the switching cost $l^s_n$ of edge server $n$ to load and evict models can be calculated as
\begin{equation}
    l^{switch}_n(\mathbf{a}^t) = \sum_{i\in \mathbb{I}} \sum_{m\in\mathcal{M}} \lambda \mathbf{1}(a_{n,i,m}^{t} > a_{n,i,m}^{t-1}),
\end{equation}
where $\lambda$ denotes the coefficient for loading and evicting the model and $\mathbf{1}(\cdot)$ is the indicator function. When $a_{n,i,m}^{t} > a_{n,i,m}^{t-1}$, i.e., $a_{n,i,m}^{t}=1$ and $a_{n,i,m}^{t-1} = 0$, $\mathbf{1}(a_{n,i,m}^{t} > a_{n,i,m}^{t-1})$ indicates that the loading of an uncached model. Otherwise, there is no switching cost incurred at edge servers. 

When the requested models are cached into the GPU memory of edge servers, users communicate with the edge servers for requesting generative AI services. Let $l_n^{trans}$ denote the transmission cost of input prompts and inference results. The transmission cost of edge server $n$ can be calculated as
\begin{equation}
    l_n^{trans}(\mathbf{a}^t, \mathbf{b}^t) = \sum_{i\in\mathcal{I}} 
    \sum_{m\in\mathcal{M}} l_{n,m} R_{n,i,m}^t a_{n,i,m}^{t} b_{n,i,m}^{t},
\end{equation}
where $r_{i, m}$ is the unit transmission cost per input and result for model $m$ of application $i$.

Let $f_n$ denote the computing capacity of edge server $n$. The forward propagation process of AI services at edge servers incurs inference latency, which can be denoted as $l^{comp}_n$ for edge server $n$. The edge computing cost can be calculated as
\begin{equation}
    l^{comp}_n(\mathbf{a}^t, \mathbf{b}^t) = \sum_{i\in\mathcal{I}} 
    \sum_{m\in\mathcal{M}} R_{n,i,m}^t a_{n,i,m}^{t} b_{n,i,m}^{t} \frac{c_n}{f_n}.
\end{equation}
Finally, as edge servers might not have sufficient resources for executing the best match model requested by AI services, the requests processed by other PFMs with the equivalent function incur accuracy cost $l^{acc}_n$, which can be represented as
\begin{equation}
    l^{acc}_n(\mathbf{a}^t, \mathbf{b}^t) = \sum_{i\in\mathcal{I}} 
    \sum_{m\in\mathcal{M}} (1-A_{i,m}) R_{n,i,m}^t a_{n,i,m}^{t} b_{n,i,m}^{t}.
\end{equation}
By sacrificing some accuracy of generative AI services, the system can reduce the model missing rate.
Therefore, the total edge inference cost of edge server $n$ is
\begin{equation}
\begin{aligned}
    L^t_n(\mathbf{a}^t, \mathbf{b}^t) = &l^{switch}_n(\mathbf{a}^t_n) + l_n^{trans}(\mathbf{a}^t_n, \mathbf{b}^t_n) \\&+ l^{comp}_n(\mathbf{a}^t_n, \mathbf{b}^t_n) + l^{acc}_n(\mathbf{a}^t_n, \mathbf{b}^t_n).
\end{aligned}
\end{equation}
The edge inference cost is jointly determined by the caching decisions and offloading decisions of edge servers. Nevertheless, the missed or offloaded requests are processed by the cloud data center.

\subsubsection{Cloud Inference Cost}

The edge servers are resource-constrained, such that they are unable to serve all PFMs. On one hand, due to the limited storage resources of the edge server, the model requested by a user may be too large to be loaded into the GPU of the edge server. On the other hand, the limited computing power of the edge server makes it necessary to actively migrate some requests to the cloud data center for execution. Therefore, when the requested models are missed at edge servers or offloaded to cloud data centers, this part of user requests are transmitted to the cloud data center, which needs to allocate resources for accomplishing such user requests. In line with~\cite{zhao2022edgeadaptor}, the cloud data centers can consider serving generative AI services in a serverless manner, which is charged in a ``pay-as-you-go" manner. Therefore, users need to pay for executing AI services according to the number of requests instead of specific occupied resources. When the requests are missed at edge servers or edge servers do not have enough resources for serving the requests, the unaccomplished requests will be offloaded to the cloud data centers for remote execution. The cloud data centers can execute the models with their abundant computing and energy resources, and then return the inference results to edge servers. However, cloud inference incurs additional latency for data transmission in the core network, which is much higher than the data transmission latency at edge servers. Moreover, the accuracy cost of offloaded inference requests executed by the cloud data center is expected to be almost zero as they can be processed by the most accurate model with common in-context examples owned by the data center. Based on the above analysis, we use $l_{0,n}$ to denote the aggregated cost of offloading one request to the cloud data center for remote execution of model $m$. Then, the total cloud computing cost at time slot $t$ is
\begin{equation}
    L^t_0(\mathbf{a}^t, \mathbf{b}^t) = \sum_{n\in\mathcal{N}\backslash\{0\}} \sum_{i\in\mathcal{I}} \sum_{m\in\mathcal{M}} l_{0,m} (1- a_{n,i,m}^t b_{n,i,m}^t) R_{n,i,m}^t.
\end{equation}



\subsection{Problem Formulation}

To optimize the performance of mobile edge intelligence, we jointly consider the cost of edge inference and cloud inference, including the switching cost, the accuracy cost, the transmission cost, and the inference cost over a time horizon $T$. The problem is formulated as follows:
\begin{mini!}|s|[2]<b>
    {\mathbf{a}^t, \mathbf{b}^t}{\frac{1}{T}\sum_{t\in\mathcal{T}}\left(L^t_0 + \sum_{n\in\mathcal{N} }L_n^t\right)\label{eq:obj}}{}{}
    \addConstraint{(1), (2), (3)}{\label{eq:con1}}
    \addConstraint{a_{n,i,m}^t\in}{ \{0,1\}\label{eq:con2}}
    \addConstraint{b_{n,i,m}^t\in}{[0,1]\label{eq:con3}}.
\end{mini!}

To solve the optimization problem described above, we must overcome the following challenges: (i) The problem involves time-coupling elements, such as GPU memories and in-context examples, as it considers both future request dynamics and historical inference contexts; (ii) Through historical statistical data, we can forecast future information before making a decision The problem becomes a mixed-integer programming problem, which is NP-hard.
To address these challenges, a low-complexity heuristic algorithm is needed to make decisions regarding model caching and request offloading, despite the lack of future information.

\section{The Least Context Algorithm}

To effectively serve PFMs for provisioning generative AI services, we propose the least context algorithm based on the AoC metric. When additional GPU memory is required for loading an uncached requested PFM, the LC algorithm counts the number of examples in context, calculates them, and removes the cached PFM with the fewest effective examples in context. Therefore, at each time slot $t$, the model caching decisions can be obtained by solving the maximization problem of the number of effective examples for the cached models, which can be represented as
\begin{maxi!}|s|[2]<b>
    {\mathbf{a}^t}{\sum_{i\in\mathcal{I}}\sum_{m\in\mathcal{M}} K_{i,m}^t \label{eq:tobj}}{}{}
    \addConstraint{\sum_{i\in \mathcal{I}} \sum_{m \in \mathcal{M}} a_{n, i, m}^t s_{m} \leq G_n^t, \forall n\in \mathcal{N}}{\label{eq:tcon1}}
    \addConstraint{a_{n,i,m}^t\in}{ \{0,1\}\label{eq:tcon2}}.
\end{maxi!}
The available capacity of GPU memory $G_n^t$ of server $n$ at time slot $t$ can be calculated as $G_n^t = G_n-R_{n,i,m}^t a_{n,i,m}^t b_{n,i,m}^t s_m$. This optimization problem can be tackled with a complexity of $O(IM)$ with prior knowledge and statistical data.
This algorithm gives the least important PFM for the current inference task priority for eviction. It works well with huge numbers of PFMs on edge servers with limited GPU memory. By using more contextual information during inference, the PFMs of mobile generative AI services are more accurate. Based on caching decisions $\mathbf{a}^t$ by solving the optimization problem (\ref{eq:tobj}), offloading decisions $\mathbf{b}^t$ are obtained by solving the optimization problem (\ref{eq:obj}).

\section{Numerical Results}

In the experiment, we consider an edge intelligence system with $T = 100$ slots. The requests for generative AI services per time slot follow the Poisson process with an average of one. We consider three types of PFMs and select six representative models to serve in the experiments, i.e., GPTs, Uniformers, and CLIPs. The detailed model configuration can be found in~\cite{xu2023sparks}. The main parameters are listed in TABLE~\ref{tab:exp}.
\begin{figure*}[t]
	\centering
	\begin{minipage}[t]{0.33\linewidth}
		\centering
		\includegraphics[width=1\linewidth]{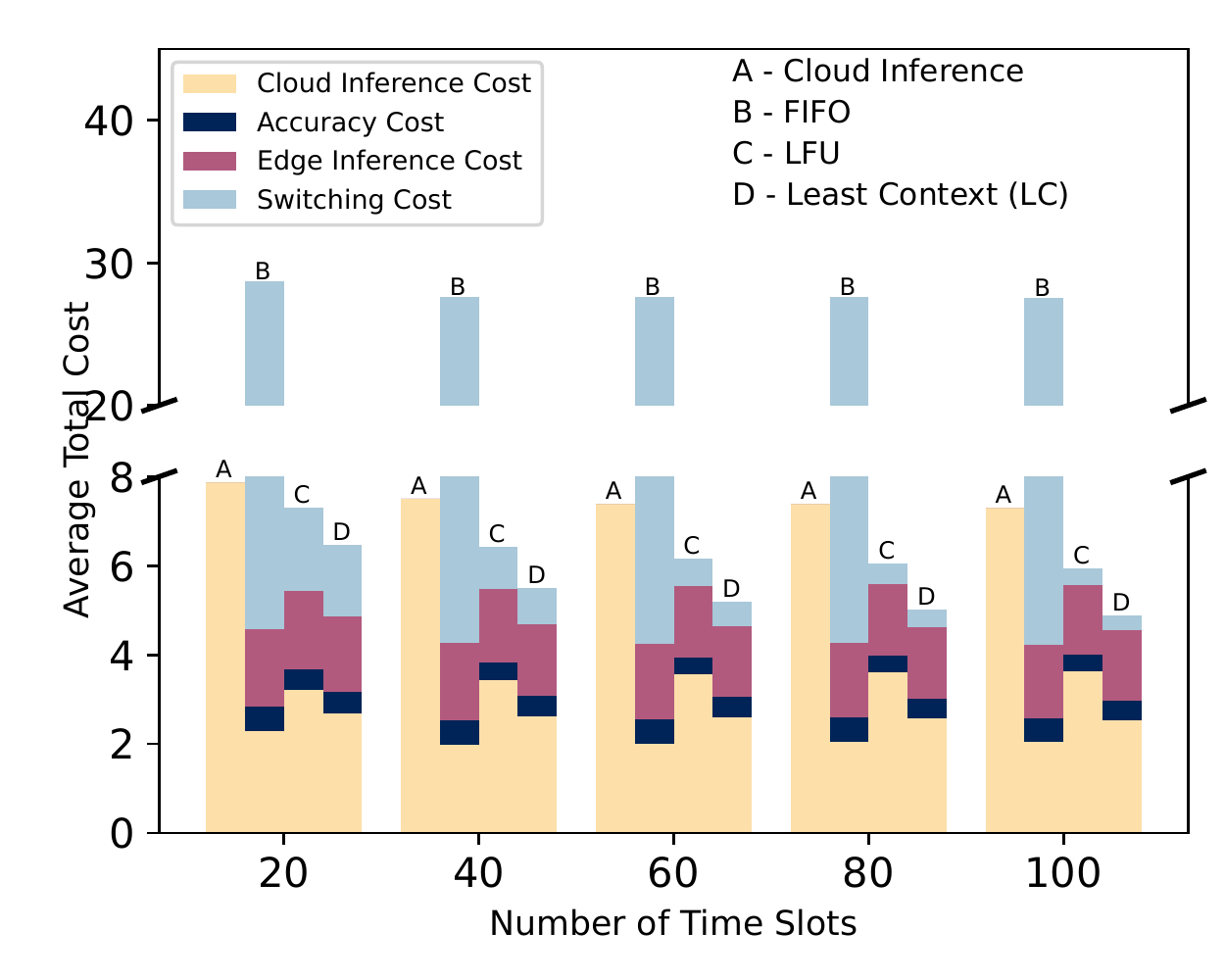}
		\caption{Average total cost versus number of time slots.}
		\label{fig:exp_T}
	\end{minipage}%
        \begin{minipage}[t]{0.33\linewidth}
		\centering
		\includegraphics[width=1\linewidth]{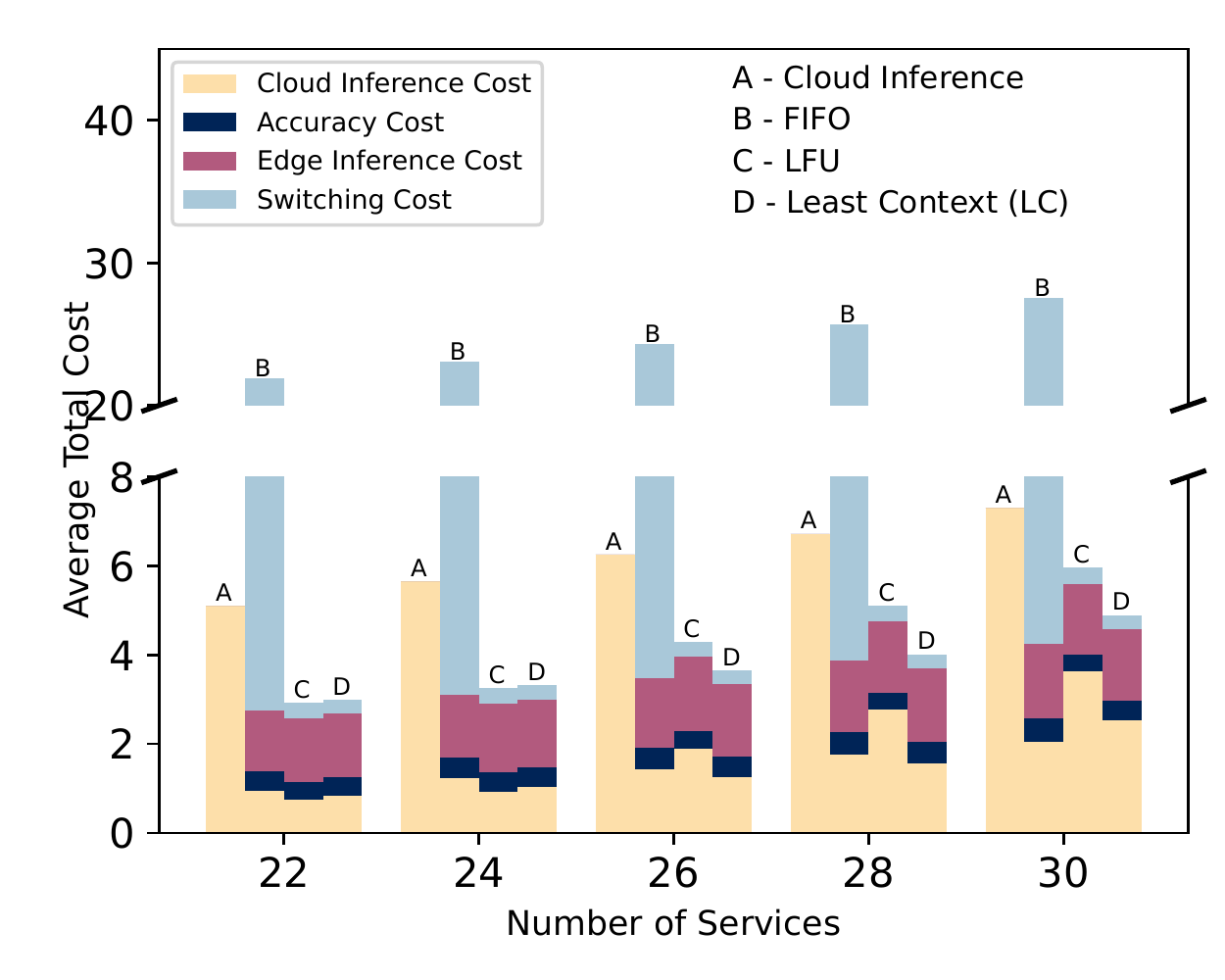}
		\caption{Average total cost versus number of services.}
		\label{fig:exp_service}
	\end{minipage}%
	\begin{minipage}[t]{0.33\linewidth}
		\centering
		\includegraphics[width=1\linewidth]{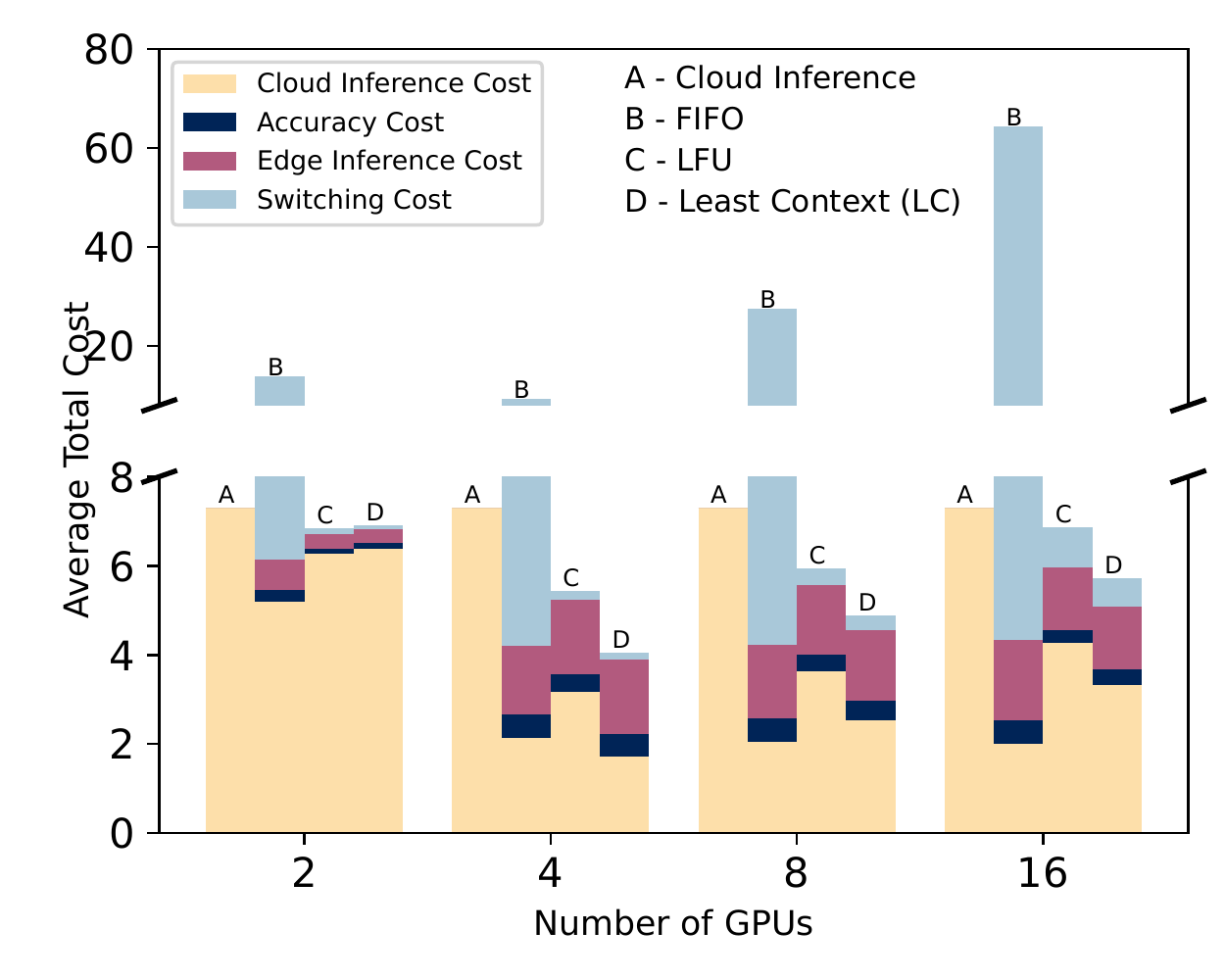}
		\caption{Average total cost versus number of GPUs.}
		\label{fig:exp_GPU}
	\end{minipage}
\end{figure*}

\begin{table}[t]
\small\centering
\caption{The List of Main System Parameters.}
\label{tab:exp}
\begin{tabular}{c|c}
\hline
Parameters                 & Value         \\
\hline
Number of time slots       & 100           \\
Number of services         & 30            \\
Number of GPUs             & 8             \\
Size of context window     & 2048          \\
Size of examples           & {[}10, 100{]} \\
GPU Memory                 & 80 GB         \\
GPU Computing Capacity     & 312000 GFLOPS    \\
Edge transmission cost     & 0.0001        \\
Cloud inference cost       & 0.0015        \\
Switching cost coefficient & 0.0001        \\
Accuracy cost coefficient  & 0.01         \\
GPU energy efficiency  & 810 GFLOPS/W         \\
Energy capacity  & 300 W         \\
\hline
\end{tabular}
\end{table}

We evaluate the proposed LC approach in comparison to several baselines including cloud inference, the first-in-first-out (FIFO) caching algorithm, and the least frequently used (LFU) caching algorithm. Initially, we examined the effectiveness of the LC algorithm by comparing the average total cost in various system settings. As we  observe in Fig.~\ref{fig:exp_T}, the switching cost of the LC algorithm gradually converges to a smaller value at around 1.3\%, while the switching cost of the FIFO algorithm remains constant with system time. This indicates that the LC algorithm is able to cache most of the required models for inference services on the edge server in GPU memory. In addition, the LC algorithm achieves the lowest average total cost among all the algorithms. The LC algorithm can reduce the cloud inference cost by increasing the utilization of edge computing resources so that the requests can be executed at edge servers with low latency.

We then show that the proposed LC algorithm is robust under different system settings, such as a different number of services and a different number of GPUs. From Fig.~\ref{fig:exp_service}, we can see that the total system cost increases with the number of services. This is because the resources in the edge servers become insufficient when more services need to be served on the edge servers. On one hand, the GPU memory on the edge servers is limited, and as the number of services increases, more model switching will be required when running the model, so the switching cost becomes higher. On the other hand, when the resources on the edge servers are not sufficient, the requests for cloud inference have to be forwarded to cloud data centers, whose costs are higher than those of edge inference. In the meanwhile, the experimental results in Fig.~\ref{fig:exp_GPU} indicate that the number of GPUs has a complex impact on the total system cost. When the number of GPUs increases, the switching cost increases. The reason is that the edge servers can cache more models in the GPU memory. Without effective management of cached models, the switching cost is high for the FIFO algorithms. Though the cost of the proposed LC algorithm is always lower than those of the other algorithms, its cost increases when the number of GPUs increases. The reason behind this trend is that edge servers can cache larger models when the number of GPUs is large. However, such large models require intensive computing resources while incurring similar edge inference costs. Therefore, these user requests for large models are better offloaded to cloud data centers for remote execution.

\begin{figure}[t]
    \centering
    \includegraphics[width=0.65\linewidth]{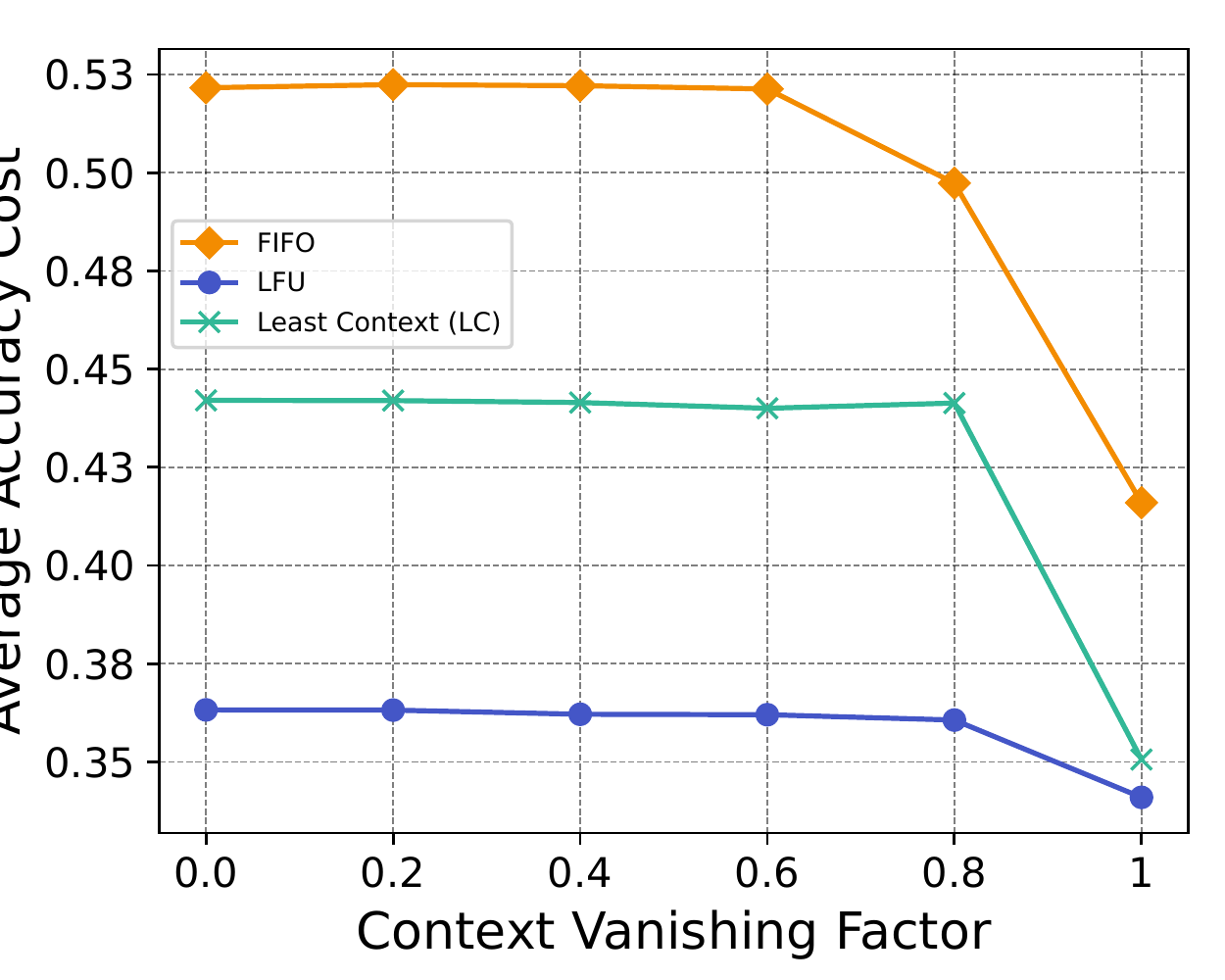}
    \caption{Edge accuracy cost vs. context vanishing factor.}
    \label{fig:acc_factor}
\end{figure}

\begin{figure}[t]
    \centering
    \includegraphics[width=0.65\linewidth]{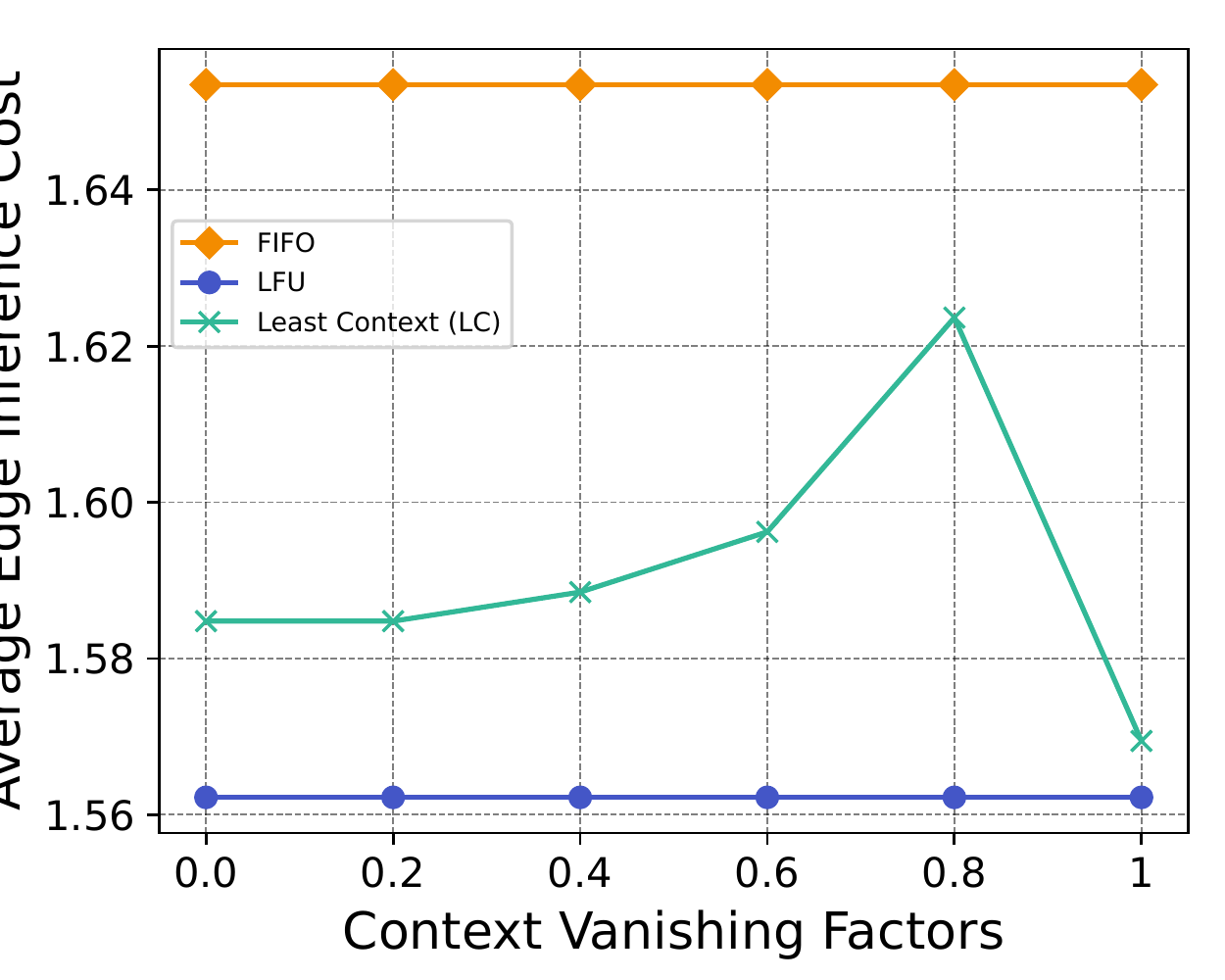}
    \caption{Edge inference cost vs. context vanishing factor.}
    \label{fig:cost_factor}
\end{figure}

After demonstrating the effectiveness of the proposed LC algorithm, we next investigate the impacts of the context vanishing factor. To make comparisons between models more noticeable, the size of the context window is set to $2^{14}$. As shown in Fig.~\ref{fig:acc_factor}, as the context vanishing factor increases, the average accuracy cost of edge inference is first static and then decreases. When the context vanishing factor is small, the performance gap among these three algorithms becomes large. However, when the context vanishing factor is large, such as one in Fig.~\ref{fig:acc_factor}, the average accuracy cost decreases, and their performance gap starts to shrink. Another interesting finding shown in Fig.~\ref{fig:cost_factor} is that the averse edge inference cost first increases as the context vanishing factor increases and then dramatically declines after a certain threshold.

\section{Conclusions}\label{sec:conclusions}

In this paper, we investigated the joint foundation model caching and inference problem for deploying PFMs to serve AI-based multimedia services in mobile edge networks. We introduced a joint foundation model caching and inference framework designed to effectively provision generative AI services at edge servers, and thus advancing toward AGI. To this end, we proposed a new metric for measuring the relevance and freshness of contextual examples in relation to ongoing inference requests. Moreover, we have developed the LC algorithm for PFM management, which optimizes the utilization of historical contextual prompts and inference results, subsequently enhancing the performance of generative AI services. Experimental results indicated that the LC algorithm effectively reduces system costs by effectively leveraging historical demonstrates and managing cached models.

\bibliographystyle{IEEEtran}
\bibliography{main}
\end{document}